\documentclass[12pt,aps,prl]{revtex4}
\usepackage{amssymb}

\usepackage{graphicx}
\input epsf

\begin{document}

\begin{center}

\
\

{\Large\textbf {Mating Instabilities Lead to Sympatric
Speciation}}

Catarina R. Almeida* \& Fern\~{a}o Vistulo de Abreu

*Depart. de Biologia, Universidade de Aveiro, 3800 Aveiro, Portugal

Depart. de F\'{i}sica, Universidade de Aveiro, 3800 Aveiro, Portugal
\end{center}
\begin{center}
\ First Version: February 2002; Revised: May 2002 \ \

\ \ {\small Corresponding author: Fern\~{a}o Vistulo de Abreu
\underline{abreu@fis.ua.pt} }
\end{center}

\strut \ \
\begin{center}
\textbf{ABSTRACT}
\end{center}

One of the most challenging issues of evolutionary biology
concerns speciation, the emergence of new species from an initial
one. The huge amount of species found in nature demands a simple
and robust mechanism. Yet, no consensus has been reached
concerning a reasonable disruptive selection mechanism that
prevents mixing genes among the emerging species, especially when
they live in sympatry. Usually it is assumed that females select
males according to their displaying traits, but males perform no
selection on female traits. However, recent experimental evidence
accumulates towards the existence of male choice. Here we propose
a robust mechanism for sympatric speciation, based on the
assumption that sexual selection operates in two directions:
selection of males by females and of females by males. Complex
mating instabilities emerge, creating differential fitness
depending on the individuals displaying traits and preferences.
When a secondary sexual trait is introduced in a population, due
to mutations, the activation of previously neutral genes or due to
a different perception of already existent displaying traits,
sympatric speciation may result (together with a species
recognition system) from a competitive exclusion principle. We
suggest that potential candidates to test our theory could be
yeasts.
\ \

\

Key index words: sympatric speciation, sexual competition, mating
systems

\strut \ \

\begin{center}
\newpage
\textbf{TEXT}
\end{center}

Traditionally, two main classes of models try to explain speciation.
Allopatric speciation models are the most consensual and assume that the
initial population is suddenly divided in two geographically isolated
subpopulations, that then diverge until they become reproductively isolated,
even after a secondary contact (Mayr 1963). There is however, mounting
evidence (Albertson \textit{et al.} 1999; Panhuis \textit{et al.} 2001;
Schliewn \textit{et al.} 1994) that speciation might have emerged in
sympatry (without geographical isolation). Examples like the cichlids in
lake Victoria or many migratory birds, do not seem to fit the basic
requirement of long periods of geographical isolation needed for allopatric
speciation. Laboratory experiments have also shown that in principle,
sympatric speciation is possible (Rice \& Hostert 1993). Understanding how
sympatric speciation can be driven, has thus attracted much theoretical
effort. However, some early works showed that finding a biologically
reasonable and robust model seems not to be an easy subject (Felsenstein
1981; Maynard Smith 1966).

Recently, several models have been proposed to explain sympatric speciation.
Nevertheless, no consensus has been reached concerning the main driving
mechanism, nor the conditions required in practice. One class of models
concentrates on sexual competition for mates. So far, these models have been
the least convincing models (Turelli \textit{et al.} 2001) as they have been
incapable of providing a robust mechanism, valid under reasonable initial
conditions (Takimoto \textit{et al.} 2000; Turelli \textit{et al.} 2001).
The other class of models assumes that competition for resources is
essential to create disruptive selection (Doebeli \& Dieckmann 2000;
Kondrashov \& Kondrashov 1999; see a discussion in van Doorn \& Weissing).
However it also requires a non-trivial association between traits relevant
for ecological adaptation and traits used for sexual discrimination.
Further, and in particular in the Doebeli and Dieckmann theory, phenotypes
using scarce resources may substitute phenotypes using more abundant
resources, as a result of a strong competition for resources in the later
case. This is a somehow counter-intuitive result. Again it remains uncertain
which species should be more speciation prone, in practice.

Here we propose an alternative approach that leads to a robust theory for
sympatric speciation. We give a special emphasis to pair formation. Pair
formation is an essential issue in any sexual population, as it is at the
basis of the species reproductive success. As we will show, modelling pair
formation is essential to create linkage disequilibrium due to the emergence
of complex mating instabilities.

In our model we assume that any individual possesses a preference list,
where all individuals of the opposite sex are ranked, in descending priority
order. The goal of any individual is to optimise his satisfaction by mating
with convenient partners. How the preference list is established depends on
a response to a sensory stimulus induced by varied secondary sexual traits
displayed by the opposite sex individuals. In principle, it could also have
an ecological component, to model, for instance, the preference for certain
habitats (Bush 1969).

As the number of displaying traits can be quite large and perceived
differently by different individuals, a simple assumption will be to
consider that each individual has his own preference list and that this is
approximately random. In this way we assume that sexual selection has
already eliminated non-favoured traits, so that only neutral traits,
concerning sexual selection, remain. Note that sexual selection does not
necessarily mean the selection of traits relevant for adaptation. Sexual
selection may only improve the fitness of a population, by optimising the
consequences of the mating conflicts created.

Sexual conflicts lead to intricate dynamics, as both males and females
pursue selfish goals, with conflicting wishes. Consider the following
preference lists, for a population with three males and three females.

\begin{center}
\strut
\begin{tabular}{|l|l|l|l|}
\hline & \textbf{M1} & \textbf{M2} & \textbf{M3} \\ \hline
\textbf{ 1 } & F1 & F2 & F2 \\ \hline \textbf{ 2 } & F2 & F3 & F1 \\
\hline \textbf{ 3 } & F3 & F1 & F3 \\ \hline
\end{tabular}
\ \ \ \
\begin{tabular}{|l|l|l|l|}
\hline \  & \textbf{F1} & \textbf{F2} & \textbf{F3} \\ \hline
\textbf{ 1 } & M2 & M1 & M2 \\ \hline \textbf{ 2 } & M1 & M3 & M3 \\
\hline \textbf{ 3 } & M3 & M2 & M1 \\ \hline
\end{tabular}
\end{center}

If all individuals search continuously for better partners, then we can
imagine the following algorithm. First, males propose sequentially to
females ranking higher in their preference lists, and females accept the new
partners only if they improve their satisfaction. Then females and males
switch roles. In this case, both males and females adopt active strategies
towards mating (they both search for better mates). We could sketch that M1
mates F1, M2 mates F2, but then F2 divorces M2 and mates with M3 because he
is higher in her preference list, and so on. It is easy to conclude that no
stable arrangement is reached. That is, the system has no Nash equilibrium
(Gale \& Shapley 1962; Omero \textit{et al.} 1997). While these
instabilities are not the rule (for instance, a couple can be stable if they
prefer each other), they produce differential costs for reproduction, as
fitness depends on the stability of the couples. There is an algorithm, due
to Gale and Shapley (Gale \& Shapley 1962; Omero \textit{et al.} 1997), that
leads to stable arrangements. This happens if males propose to females, and
females dispose, accepting to mate new pretenders only if they improve their
satisfaction. Nevertheless, even if a stable solution exists, the time
needed to reach it, tGS, tends to be quite large, due to the large number of
degrees of freedom and the complex dynamics. In practice, the results we
will present do not depend on the strategy adopted by females towards mating.

\strut

\ \

\begin{center}
\textbf{Methods}
\end{center}

We consider discrete generations of a population with an equal number, $%
N_{tot}$, of males and females. The genetic information on each individual
controls several characters. A sexual locus may control the strategy towards
mating if one would want to study the selection of these strategies, but
this, we checked, does not change the results concerning speciation.
Henceforth males and females are assumed to take active and passive
strategies towards mating, respectively.

One locus with two alleles (we assume dominance) accounts for a secondary
sexual trait (\textit{A} or \textit{B}). Several independent loci with two
alleles ($+$ or $-$) define a quantitative preference for one trait. Each
individual ranks in a list of preferences individuals of the opposite sex.
Preference lists are defined with a simple step-like structure. Calling $%
n_{+}$ the fraction of positive alleles, then a positively (negatively)
biased arrangement corresponds to a preference for mates with trait \textit{A%
} (\textit{B}), such that there is a fraction $f_{\nu
}(n_{+})=[4n_{+}(1-n_{+})]^{\nu }/2$ of non-preferred mates ranked within
the first $N_{A}$ ($N_{B}$) positions ($N_{A}$ and $N_{B}$ being the number
of individuals respectively with phenotype \textit{A} and \textit{B}). The
positive exponent n controls how efficiently individuals with a biased
arrangement of alleles discriminate mates with preferred traits. For small
values of $\nu $, deviations from an unbiased arrangement produces poor
discrimination, as can be seen in figure 1.

On each iteration, randomly picked individuals (with an active strategy
towards mating) are given one opportunity to find a better partner. New
matings happen only if they improve the satisfaction of both elements of the
couple. The process repeats until $2N_{tot}$ offspring are born. One
offspring of each sex is born if a couple stays together during a courtship
time tc, plus a reproduction time tr (results concerning the emergence of
speciation do not change if offsprings are only allowed to born after a
stabilization time). During reproduction time, females cannot engage in new
matings, but males may find a better partner (in relation to speciation, the
relevant parameter is simply $t_{c}+t_{r}$). Offspring genotypes are found
using Mendel genetic rules.

By introducing a probability m of changing each allele controlling the trait
of the offspring, we can model phenomenologically a many genes dependence on
the trait or the effect of mutations. A similar procedure can be applied to
the loci controlling the preference.

Finally, ornamentation costs can be introduced within our framework by
assigning a birth probability, related to the parent's trait. We checked
that speciation is not prevented if the ornamentation costs are not too
different, and thus this point will not be discussed further here.

\strut \ \

\begin{center}
\textbf{Results and Discussion}
\end{center}

Consider first a population with only one trait. In figure 2, the
distribution function of mating lifetimes is shown. It exhibits clear power
law decay. This scale invariant behaviour means that there is a large amount
of stable matings, but also a large amount of very unstable short-lived
matings, relatively to what one would expect if matings were random. Thus
pair formation will certainly play a role in a sexual selection theory, as
it distinguishes individual fitnesses over several scales. In what concerns
speciation, another important consequence is that couples stability is
context dependent: the same couple may have very different lifetimes
depending on the preferences present in the whole population.

Note that scale invariant laws are often associated to self-organized
criticality (Jensen 1998). Here, however, it results from the existence of
preference lists. It is nevertheless conceivable that other alternative
rules exist leading to the emergence of a similar scale invariant behaviour
in a self-organized way.

For speciation to emerge as a result of sexual selection, a new sexual
secondary trait must be introduced in a population (due to mutations, or to
the activation of certain genes (Rutherford \& Lindquist 1998), induced by
new ecological conditions), or already existent displaying traits should be
perceived differently under new ecological conditions (for instance, water
quality or light intensity in a lake (Boughmann 2001)), differentiating the
individuals. Thus, a population subject to new ecological conditions may be
driven into a sexual selection process, from which, we will show, sympatric
speciation can be the outcome.

Consider the limit case in which an individual would rank all the mates with
the preferred trait first. In general, if a new trait (\textit{A} or \textit{%
B}) is introduced in a population, some individuals may prefer a mate with
trait \textit{A}, while others, one with trait \textit{B}. We can classify
males and females in four groups, \textit{AA}, \textit{AB}, \textit{BA} and
\textit{BB}, the first letter standing for the trait, the second for the
preference. If only males adopt active strategies towards mating, then both,
\textit{AA} and \textit{BA} males, compete for the same females, \textit{AA}
or \textit{AB}. A potential opportunity cost exists if, for instance, a
\textit{BA} male mates with an \textit{AA} female, as there is a high
probability that this mating will not last enough time to accomplish
reproduction. We have checked that the mating instabilities reduce strongly
the lifetime of couples where traits and preferences do not match. From the
female's point of view, \textit{AA} and \textit{BA} females prefer \textit{AA%
} or \textit{AB} males, which produces an indirect competition between
\textit{AA} and \textit{AB} males. Neglecting the effects of recombination
and assuming that the average number of males $ij$, $m_{ij}$, is equal to
the average number of females with matched preferences, $f_{ji}$ , then $%
m_{ij}$ grows roughly with the probability of forming stable couples, $%
dm_{ij}\simeq \alpha \;m_{ij}^{2}/m_{j},$ where $m_{j}$ is the total number
of males with preference for $j$. Sexual selection will only keep those
individuals reproducing at the highest rate, leading to a \emph{competitive
exclusion principle} with a symmetry breaking, where only \textit{AA}$\times
$\textit{AA }and \textit{BB}$\times $\textit{BB }couples or \textit{BA}$%
\times $\textit{AB }and \textit{AB}$\times $\textit{BA}, are selected.
Introducing recombination with one dominant locus for each, trait and
preference, this competitive scenario can be tackled analytically as shown
in the appendix. Then, only \textit{AA}$\times $\textit{AA }and \textit{BB}$%
\times $\textit{BB }couples are selected, which corresponds to the emergence
of homotypic preferences.

Our general model considers a quantitative genetic preference, determined by
several additive independent loci. The number of these loci does not need to
be only one, as considered in the simple case in the appendix. If we
consider a large number of loci controlling the preferences, then random
configurations produce only small preference tendencies. The more biased
arrangements of preference alleles are, the larger will be the number of
individuals with the preferred trait ranked in the first positions. As
discussed above, the function $f_{\nu }(n_{+})$, establishes the relation
between bias in preference loci and the intensity of discrimination. In
figure 3a we show how sympatric speciation emerges, starting from a
population with random genotypes. Even for species with a considerable
number of independent genes controlling the preferences, speciation is
remarkably achieved within just a few generations. Due to trait dominance,
only \textit{AA} or \textit{BB} individuals last in the end, all
heterozygous being eliminated. This corresponds to the emergence of two new
species and a species recognition system. Prezygotic isolation is thus
achieved giving way to species divergence. In this example, trait
discrimination is the same for males and females (same $\nu $). However, for
this example speciation could also be achieved if we had chosen higher trait
discrimination for females (higher $\nu $, $\nu _{females}=1.5$ and $\nu
_{males}=0.5$) as it is likely to happen in nature. In fact, trait
discrimination needs to operate on both sexes, but its intensity may vary.

If $f_{\nu }(n_{+})$ departs slowly from the unbiased arrangement of alleles
corresponding to a difficult discrimination, disruptive sexual selection can
be prevented (figure 3b). It is also possible to prevent speciation if
traits are not straightfully inherited, due to complex dependencies on a
large number of other genes (a trait may result from intricate interactions
of a large number of proteins, leading to a complex inheritance) or as a
result of mutations. These effects are taken into account through the m
parameter. Large values of $\mu $ (for the example of figure 3a, $\mu
\approx 6\times 10^{-3}$ mutations per generation and per allele) reduce the
effects of the non-random pair formation created by the complex dynamics.
Consequently above a certain threshold value speciation is prevented and a
population with genotype frequencies in the Hardy-Weinberg proportions
remains.

Our theory assumes that in a population, each individual presents many
phenotypic traits, each contributing in a complicated way to establish the
other individuals mate choice. Many of these traits may not have a strong
impact in mate choice, in such a way that a phenotypic polymorphism remains
in a population (they would be approximately neutral in what concerns sexual
selection). In figure 4a we show a population in which discrimination
relatively to a given trait is poor in the first 200 generations. The
population consists of individuals with both traits and small preferences
relatively to both traits. There is no relevant correlation between traits
and preferences in the individuals. At generation 200, discrimination for
the trait is increased and a quick speciation event emerges. An alternative
picture is presented in figure 4b. Starting from a monomorphic population
and sufficiently strong trait discrimination, mutations and a release of
natural selection pressure (as a result of change of habitat, for instance)
allows genetic drift. Preferences acquire variability, until individuals
with preferences for both traits exist in the population. Then sympatric
speciation emerges. In practice this later scenario needs more generations
to reach the speciation disruptive configuration.

The theory presented here explains, under minimal assumptions, how sympatric
speciation emerges or not, together with a species recognition system. It is
a robust mechanism, not requiring special assumptions on the initial species
(Doebeli \& Dieckmann 2000; Kondrashov \& Kondrashov 1999; Lande 1982;
Turner \& Burrows 1995; van Doorn \& Weissing), on male-female
incompatibilities (Gavrilets 2000; Parker \& Partridge 1998), on the
ecological conditions (Turner \& Burrows 1995), or on non obvious
connections between sexual and ecological traits (Doebeli \& Dieckmann 2000;
Felsenstein 1981; Maynard Smith 1966; Turelli \textit{et al.} 2001; van
Doorn \& Weissing). Our theory assumes that males and females have
preferences over the same traits. While female preferences have been
extensively studied (Andersson 1994; Futuyma 1998), male preferences are
usually neglected. Nevertheless, there is now mounting experimental evidence
for the existence of male choice (Andersson 1994; Burley \textit{et al.}
1982; Rutowski 1982; Roulin \textit{et al.} 2000; Katvala \& Kaitala 2001;
Amundsen \& Forsgren 2001). This is actually a natural outcome of natural
selection as males would otherwise incur in potential opportunity costs:
those having the ability to elicit healthier females, or predators mimicry,
should certainly be favoured. However, in practice, they should be more
difficult to detect, as males usually adopt active strategies towards
mating, and thus are seen performing courtship to many females. This is
certainly an issue deserving more research.

A unified understanding of sexual selection, starting from simple organisms
like yeasts to complex organisms like mammals, may be attempted with our
model. Indeed, in yeasts it is now clear that both mates in a couple choose
their partners (Jackson \& Hartwell 1990). Further, many genes can be
involved in the several complex steps of this process (White \& Rose 2001).
Hence, analogues to the preference lists used in our model may already be
built in. For this reason we propose yeasts as potential candidates to test
our theory in the laboratory. And if, as predicted, robust sympatric
speciation is observed, then many implications can result not only on a
conceptual level. Indeed we believe that sympatric speciation could be used
as a general framework for studying the interactions between receptors and
pheromones, and to construct a better understanding on how these correlate
with the intracellular state. This issue will be addressed in a forthcoming
publication.

\begin{center}
\strut

\textbf{ACKNOWLEDGMENTS}
\end{center}

We acknowledge Francisco Dion\'{i}sio and Benoit Dou\c{c}ot for critical
reading of the manuscript. We thank R.M.Conde for earlier collaboration, S.
van Doorn, A. Luis, M. A. Santos and I. Pimentel for discussions.

\strut

\strut

\begin{center}
\textbf{Appendix: Mathematical approach of the sexual competitive scenario
for a population with two locus, for traits and preferences, with a dominant
allele each.}
\end{center}

Here we outline a mathematical approach to the emergence of an exclusion
principle from our results. The dynamical correlations in the model reduce
strongly the fitness of non-optimal couples. Assume then that fitness is
proportional to the probability of forming couples where traits and
preferences match. In a two-locus, two-allele model, we have 9 genotypes,
denoted by $ij$, respectively associated with displaying traits and
preferences, with $i,j=1,2,3$. Genotype $2$ is heterozygous, and $1$ is a
dominant homozygous. Denote the number of males (females) with genotype $ij$
by $m_{ij}\;(f_{ij})$. The evolution of the number of genotypes $ij$ is
proportional to: $dm_{ij}\simeq \sum_{klmn}\alpha
_{ij}^{klmn}m_{kl}\,f_{mn}/f_{M}$, where $f_{M}$ is the number of females
with trait corresponding to genotype mn $(M=A,B)$ and $\alpha _{ij}^{klmn}$
is a matrix giving the reproduction frequencies following Mendel rules. This
equation is quite general but simplifies considerably in the present case,
as most entries in the matrix are zero (see below). For instance, the
equation associated with phenotype \textit{BB} (genotype $m_{33}$) is: $%
dm_{33}$ $\simeq
m_{33}\,f_{33}\,/f_{B}+m_{23}\,f_{32}\,/(4f_{B})+m_{32}\,f_{23}%
\,/(4f_{B})+m_{22}\,f_{22}\,/(16f_{A}).$

An exclusion principle emerges as genotypes try to grow the fastest possible
(following similar growth equations) while interacting and as we concentrate
on populations with a fixed number of individuals. Only those growing at the
highest rate will remain.

If we sum all genotypes contributing to the same phenotype, we arrive to
similar equations to the evolution of the phenotypes. It is possible to
check that all terms contributing to $dm_{\mathit{AB}}=dm_{13}+dm_{23}$ and $%
dm_{\mathit{BA}}=dm_{31}+dm_{32}$ appear in $dm_{\mathit{AA}%
}=dm_{11}+dm_{12}+dm_{21}+dm_{22}$ but with larger coefficients in the
later. Thus $m_{\mathit{AB}}$ and $m_{\mathit{BA}}$ are eliminated (unless
the starting population had only genotypes $13$ and $31$). Phenotypes $m_{%
\mathit{AA}}$ and $m_{\mathit{BB}}$ can coexist as both have a non-mixed
contribution corresponding to the union of genotypes $m_{11}+f_{11}$ and $%
m_{33}+f_{33}$.

The non-zero matrix entries are: ($\alpha _{ij}^{klmn}=$ $\alpha
_{ij}^{mnkl} $)

$1/2=\alpha _{111121}=\alpha _{111211}=\alpha _{121112}=\alpha
_{121212}=\alpha _{211121}=\alpha _{212121}$

$1/4=\alpha _{111122}=\alpha _{111212}=\alpha _{111221}=\alpha
_{112121}=\alpha _{121122}=\alpha _{121221}=\alpha _{121222}=$

$=\alpha _{131212}=\alpha _{211122}=\alpha _{211221}=\alpha _{212122}=\alpha
_{221122}=\alpha _{221221}=\alpha _{221222}=$

$=\alpha _{222122}=\alpha _{222222}=\alpha _{222332}=\alpha _{232332}=\alpha
_{312121}=\alpha _{322332}=\alpha _{332332}$

$1/8=\alpha _{111222}=\alpha _{112122}=\alpha _{122122}=\alpha
_{122222}=\alpha _{131222}=\alpha _{211222}=\alpha _{212222}=\alpha
_{231222}=$

$=\alpha _{232222}=\alpha _{312122}=\alpha _{322122}=\alpha _{322222}$

$1/16=\alpha _{112222}=\alpha _{132222}=\alpha _{312222}=\alpha _{332222}$

$1=\alpha _{111111}=\alpha _{221331}=\alpha _{333333}$

\strut

\begin{center}
\textbf{REFERENCES}
\end{center}

Albertson, R.C., Markert, J.A., Danley, P.D. \& Kocher, T.D. 1999 Phylogeny
of a rapidly evolving clade: the cichlid fishes of Lake Malawi, East Africa.
Proc. Natl. Acad. Sci. USA 96, 5107-5110.

Amundsen, T. \& Forsgren, E. 2001 Male mate choice selects for female
coloration in a fish. Proc. Natl. Acad. Sci. USA 98, 13155-13160.

Andersson, M. 1994 Sexual Selection. Princeton University Press.

Boughmann, J. W. 2001 Divergent sexual selection enhances reproductive
isolation in sticklebacks. Nature 411, 944-948.

Burley, N., Krantzberg, G. \& Raduran, P. 1982 Influence of colour-banding
on the conspecific preferences of zebra finches. Anim. Behav. 27, 686-698.

Bush, G. L. 1969 Sympatric host race formation and speciation in frugivorous
flies of the genus Rhagoletis (Diptera, Tephritidae). Evolution 23, 237-251.

Doebeli, M. \& Dieckmann, U. 2000 Evolutionary branching and sympatric
speciation caused by different types of ecological interactions. Am. Nat.
156, S77-S101.

Felsenstein, J. 1981 Skepticism towards Santa Rosalia, or why are there so
few kinds of animals? Evolution 35, 124-138.

Futuyma, D.J. 1998 Evolutionary Biology. Sunderland: Sinauer.

Gale, D. \& Shapley, L. S. 1962 College admission and the stability of
marriage. Am. Math. Monthly 69, 9-14.

Gavrilets, S. 2000 Rapid evolution of reproductive barriers driven by sexual
conflict. Nature 403, 886-889.

Jackson, C. L. \& Hartwell, L. H. 1990 Courtship in S. cerevisae: both cell
types choose mating partners by responding to the strongest pheromone
signal. Cell 63, 1039-1051.

Jensen, H. J. 1998 Self-Organized Criticality. Cambridge Univ. Press.

Katvala, M. \& Kaitala, A. 2001 Male choice for current female fecundity in
a polyandrous egg-carrying bug. Anim. Behav. 61, 133-137.

Kondrashov, A. S. \& Kondrashov, F. A. 1999 Interactions among quantitative
traits in the course of sympatric speciation. Nature 400, 351-354.

Lande, R. 1982 Rapid origin of sexual isolation and character divergence in
a cline. Evolution 36, 213-223.

Maynard Smith, J. 1966 Sympatric speciation. Am. Nat. 916, 637-650.

Mayr, E. 1963 Animal Species and Evolution. Harvard University Press.

Omero, M. J., Dzierzawa, M., Marsili, M. \& Zhang, Y. - C. 1997 Scaling
behavior in the stable marriage problem. J. Phys.I France 7, 1723-1729.

Panhuis, T. M., Butlin, R., Zuk, M. \& Tregenza, T. 2001 Sexual selection
and speciation. Trends Ecol. Evol. 16, 364-371.

Parker, G. A. \& Partridge, L. 1998 Sexual conflict and speciation. Phil.
Trans. R. Soc. London B 353, 261-274.

Rice, W. R. \& Hostert, E. E. 1993 Perspective: Laboratory experiments on
speciation: what have we learned in forty years?. Evolution 47, 1637-1653.

Roulin A, Jungi TW, Pfister H, Dijkstra C. 2000 Female barn owls (Tyto alba)
advertise good genes. Proc R Soc Lond B 267, 937-941

Rutherford, S. L. \& Lindquist, S. 1998 Hsp90 as a capacitor for
morphological evolution. Nature 396, 336-342.

Rutowski, D.L. 1982 Epigamic selection by males as evidenced by courtship
partner preferences in the checkered white butterfly (Pieris protodice).
Anim. Behav. 30, 108-112.

Schliewen, U. K., Tautz, D. \& Paabo, S. 1994 Sympatric speciation suggested
by monophyly of crater lake cichlids. Nature 368, 629-632.

Takimoto, G., Higashi, M. \& Yamamura, N. 2000 A deterministic genetic model
for sympatric speciation by sexual selection. Evolution 54, 1870-1881.

Turelli, M., Barton, N. H. \& Coyne, J. A. 2001 Theory and speciation.
Trends Ecol. Evol. 16, 330-343.

Turner, G. F. \& Burrows, M. T. 1995 A model of sympatric speciation by
sexual selection. Proc. R. Soc. Lond. B 260, 287-292.

van Doorn, G. S. \& Weissing, F. J. Ecological versus sexual selection
models of sympatric speciation. Selection in press.

White, J.M. \& Rose, M.D. 2001 Yeast mating: getting close to membrane
merger. Curr. Biol. 11, R16-R20.

\strut \ \

\strut \ \
\newpage

\strut \ \
\begin{figure}[h]
 \includegraphics[width=3in]{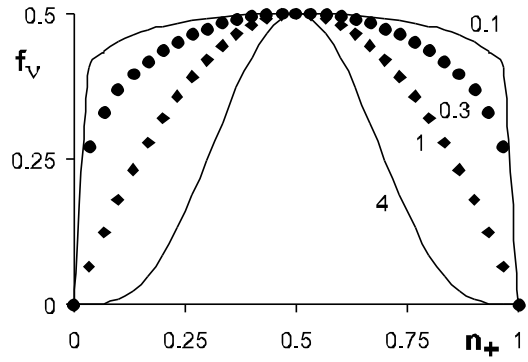}
 \caption{Preference lists are constructed with a step-like
structure. In the first positions there is a fraction $f(n_{+})$
of individuals with non-preferred traits. The parameter n controls
how efficiently individuals with a biased arrangement of alleles
($n_{+}$ away from $1/2$) discriminate mates with preferred
traits. The function $f_{\nu }$ is shown for $\nu =0.1,0.3(\bullet
),1(\blacklozenge )$ and $4$. Dots and diamonds show the discrete
values of $f_{\nu }$ for the results of figure 3 (preferences with
30 alleles).}
\end{figure}

\strut \ \
\begin{figure} [h]
      \includegraphics[width=3in,clip]{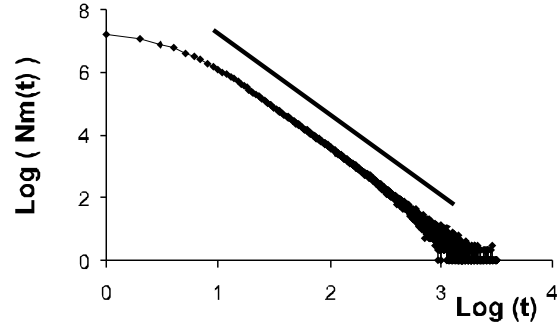}\\
 \caption{ Number of matings $N_{m}(t)$ lasting for $t$
iterations. 500 populations were sampled over 10000 iterations,
for the model without reproduction events ($t_{c}=t_{r}=+\infty
$). The distribution shows power law behaviour,
$N_{m}(t)\;\symbol{126}\;t^{-\tau }$, with $\tau \simeq 2.5$.
$\tau $ is independent of the number of individuals and the
strategies of females. This distribution shows that, due to the
correlated dynamics, individuals are expected to have fitness
varying over several orders of magnitude. Thus correlated dynamics
must play an essential role on the evolution of the population.}
\end{figure}
\strut \ \
\begin{figure}[h]
       \includegraphics[ width=6.5in]{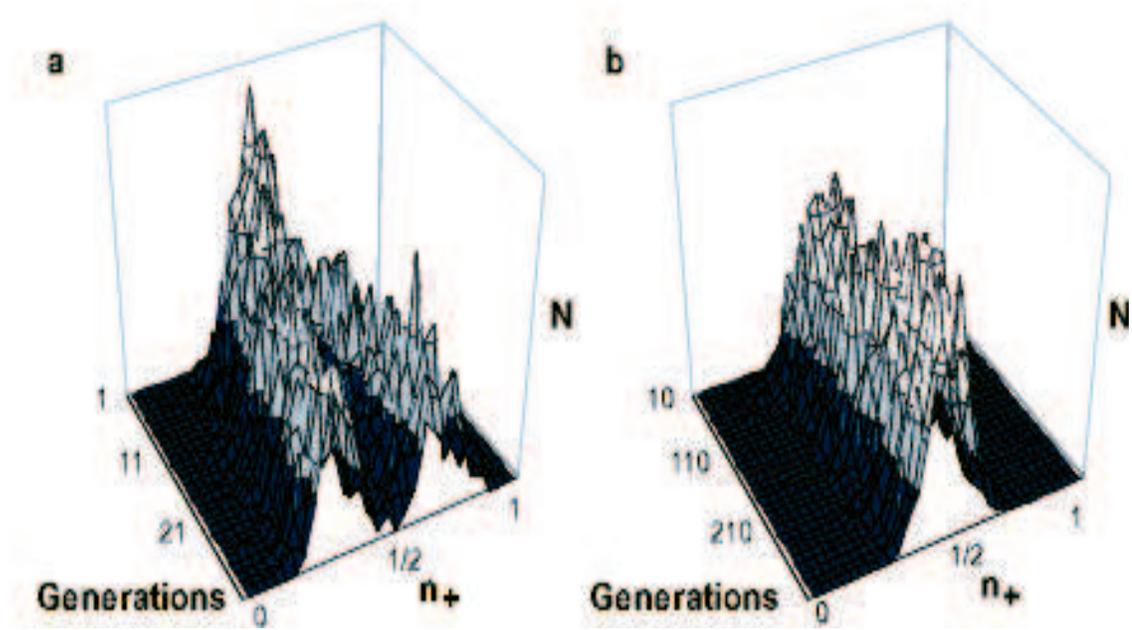}
    \caption{Evolution of phenotype frequencies
starting from a population with random initial genotypes. One
dominant locus with two alleles controls the trait, and 15
independent loci with two alleles each, the preference intensity.
For $n_{+}$ close to $1(0)$, mates with trait \textit{A}
(\textit{B}) tend to be preferred. Sympatric speciation \textbf{a}
quickly emerges if sufficient discrimination of mating traits
exists ($\nu =1.0$), or \textbf{b} is prevented if discrimination
is insufficient$\,(\nu =0.3)$ $(N_{tot}=200,t_{c}=t_{r}=40)$.}

\end{figure}

\strut \ \
\begin{figure}[h]
       \includegraphics[width=6.5in,clip]{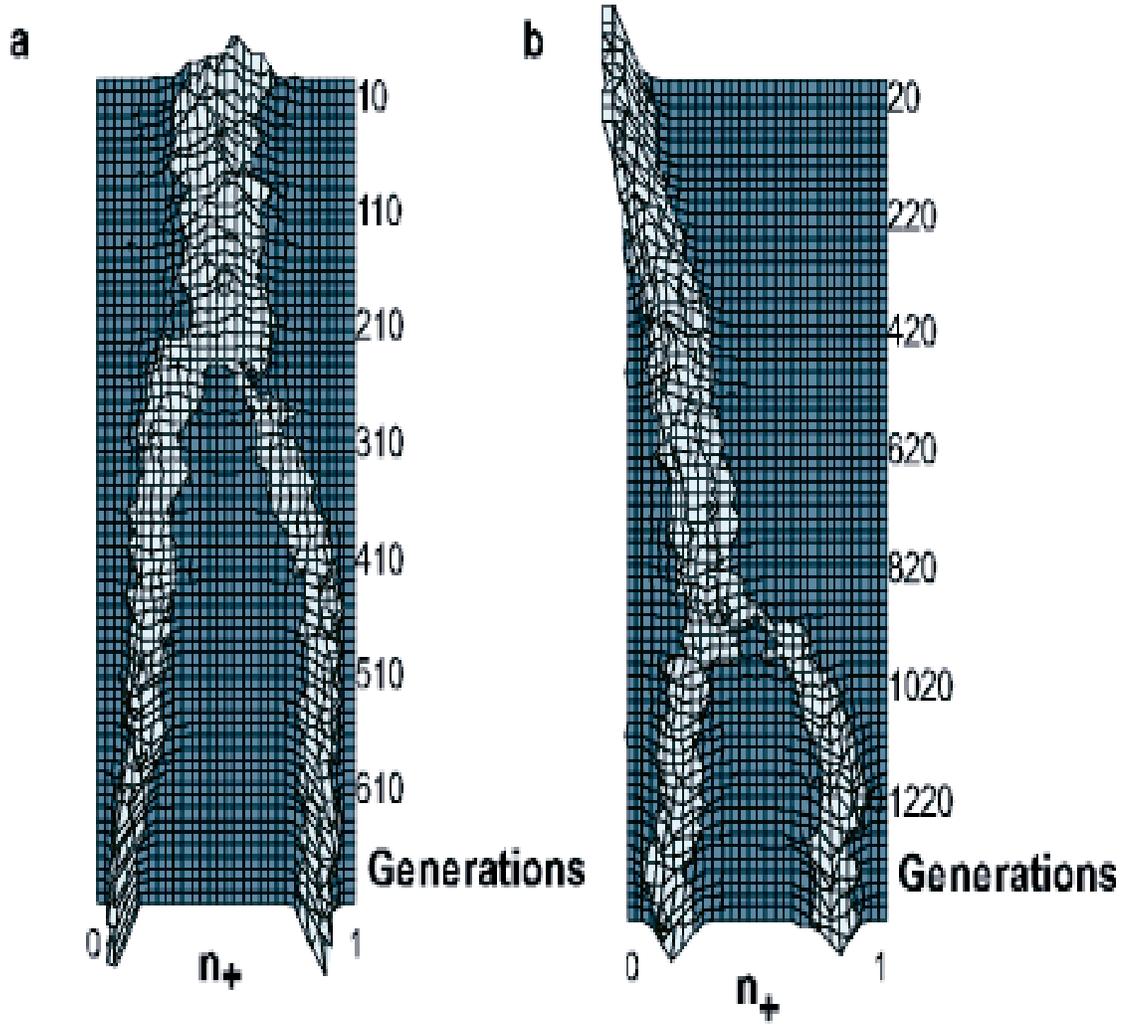}
 \caption{Evolution of phenotype frequencies
(one dominant locus controls the trait; 15 independent loci, the
preference intensity):\textbf{\ a} starting from a population with
random initial genotypes, $\mu =3\times 10^{-4}$ per allele, and
weak trait discrimination ($\nu =0.4$). At generation 200
discrimination is increased ($\nu =1.0$). \textbf{b} starting from
a monomorphic population (all individuals with trait \textit{B}
and all their preference loci equal to 0 - preference for
\textit{B} traits) and discrimination $\nu =1.0$. Mutations ($\mu
=6\times 10^{-4}$ per allele) produce genetic drift such that
speciation may emerge when there are individuals with preferences
for both traits. Light squares correspond to phenotypes with more
than 10 individuals.}
\end{figure}

\strut


\end{document}